%% file: templateArxiv.tex
\documentclass{article}

\usepackage{PRIMEarxiv}

\usepackage[utf8]{inputenc} 
\usepackage[T1]{fontenc}    
\usepackage{hyperref}       
\usepackage{url}            
\usepackage{booktabs}       
\usepackage{amsfonts}       
\usepackage{nicefrac}       
\usepackage{microtype}      
\usepackage{lipsum}
\usepackage{fancyhdr}       
\usepackage{graphicx}       
\usepackage{savesym}
\usepackage{amsmath}
\usepackage{cases} 
\usepackage{textcomp}
\usepackage{xspace}
\usepackage{verbatim}
\usepackage{soul}
\usepackage{xcolor}
\usepackage{blindtext}
\usepackage{verbatim}
\usepackage{algorithm}
\savesymbol{AND}    
\usepackage{algorithmic}
\usepackage{tabularx} 
\usepackage{enumerate}
\usepackage{multirow}
\usepackage{diagbox}
\usepackage{colortbl}
\usepackage{hhline}
\usepackage{comment}
\usepackage{balance}

\usepackage{dsfont}
\usepackage{wrapfig}
\usepackage[defaultlines=4,all]{nowidow}
\definecolor{darkblue}{rgb}{0,0,0.5} 
\usepackage{transparent}



\makeatother

\newcommand{\additiveS}{\textit{Additive}\xspace}

\newcommand{\CPSs}{CPSs\xspace}

\newcommand{\BO}{BO\xspace}

\newcommand{\turbo}{TuRBO\xspace}

\newcommand{\SUT}{SUT\xspace}

\newcommand{\TS}{TS\xspace}

\newcommand{\tauZero}[1]{\ensuremath{\tau = 0}\xspace} 
\newcommand{\tauOne}[1]{\ensuremath{\tau = -1}\xspace} 
\newcommand{\piBO}[1]{\ensuremath{\pi}BO\xspace}

\usepackage{todonotes}

\newcommand{\period}{{\it period}\xspace}

\newcommand{\delay}{{\it delay}\xspace}

\newcommand{\AT}{{\it AT}\xspace}
\newcommand{\CC}{{\it CC}\xspace}

\newcommand{\switchedS}{{\it SS}\xspace}
\newcommand{\NN}{{\it NN}\xspace}
\newcommand{\AFC}{{\it AFC}\xspace}
\newcommand{\SC}{{\it SC}\xspace}

\DeclareUnicodeCharacter{2212}{-}

\newcommand{\lowPrime}{{\it low^\prime}\xspace}
\newcommand{\highPrime}{{\it high^\prime}\xspace}
\newcommand{\periodPrime}{{\it period^\prime}\xspace}
\newcommand{\widthPrime}{{\it width^\prime}\xspace}
\newcommand{\delayPrime}{{\it delay^\prime}\xspace}

\title{Technical Report: The effect of Input Parameters on Falsification of Cyber-Physical Systems
}

\author{
  Zahra~Ramezani \\
  Chalmers University of Technology \\
  Gothenburg, Sweden \\
  \texttt{rzahra@chalmers.se} \\
  \And
  Knut {\AA}kesson \\
  Chalmers University of Technology \\
  Gothenburg, Sweden \\
  \texttt{knut@chalmers.se} \\
}

\begin{document}
\maketitle

\section{Introduction}

The aim of this technical report is to investigate the effect of input parameters on the falsification of cyber-physical systems (\CPSs). Falsification is a practical testing method that can be used to improve confidence in the correctness of the system. During the falsification process, only the input-output behavior of the system under test (SUT) can be observed. The falsification process only requires that the \SUT can be simulated and formal specifications exist. By associating a quantitative semantics with a formal specification, it becomes possible to formulate the falsification problem as an optimization problem.
Thus, reducing the number of simulations necessary to falsify a specification is an important goal. 

As optimization-based approaches are sensitive to dimensionality, the number of optimization variables should be kept to a minimum. 
Typically two parameters are required to generate an input for the optimization process for falsification; control points and the interpolation between them. On the other hand, the input signals like pulse generators need different parameters to be generated like period and amplitude.
It is often challenging to define suitable input parameters since the dynamics of the systems are complex and often unknown, particularly for large-scale systems.
Falsification can be successful or unsuccessful depending on the choice of the number of input parameters. Limiting the number of control points can make solving the problem easier by decreasing its dimensionality. The constraint on the inputs, however, might make the problem impossible to solve where no counterexample will be found. Thus, it is necessary to find a balance between the flexibility of input generation and the complexity of the optimization problem. This report investigates the effects of dimensionality on optimization problems of the falsification process through an experimental evaluation.  

This report is organized as follows: after the introduction that expressed the aim of this report, Section~\ref{Sec:Falsification} introduces the falsification process briefly. \ref{Sec:Pulse} introduces 
The evaluation is done on the benchmark examples discussed in Section~\ref{Sec:Evaluation-Results}. Finally, Section~\ref{Sec:Conclusion} summarizes the aim of this report.

\section{Falsification of Cyber-Physical Systems}
\label{Sec:Falsification}
\begin{figure} [htp]
    \centering
    \includegraphics[width=\textwidth]{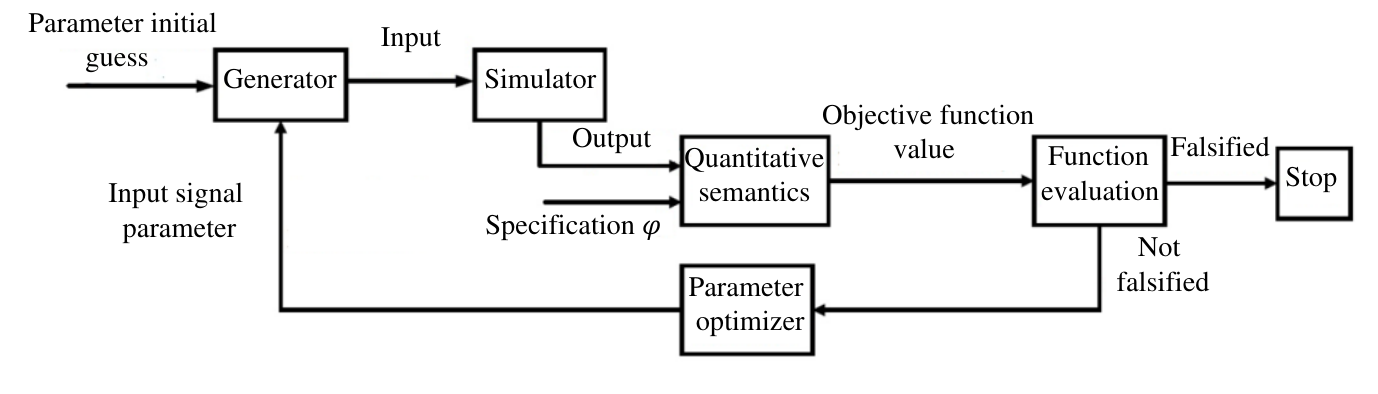}
    \caption{A flowchart of optimization-based falsification.}
    \label{fig:falsification}
\end{figure}

The process of simulation-based falsification is shown in Fig.~\ref{fig:falsification}. Initially, a generator creates input signals to the system based on an input parametrization. Next, a simulator generates simulation traces of output signals where the \SUT is simulated with the input signals. The combination of both input and output signals is used with the specification $\varphi$, possibly containing temporal operators, to evaluate the specification using a quantitative semantics. A quantitative semantics assigns an objective function value to determine whether the specification is satisfied and to what extent a specification is fulfilled. If the specification is not falsified, a parameter optimizer generates a new set of parameters for the input generator, and a new simulation of the system takes place. On the other hand, if the specification is falsified, the process ends. In this paper, signal temporal logic (STL)~\cite{Donze2010b} specifications are used to express specifications using Breach~\cite{Donze2010} as the testing tool. For quantitative semantics, \additiveS~\cite{Claessen2018, eddel2019enhancing} is used.

\section{Pulse Generator}
\label{Sec:Pulse}

Each input to the system is defined in a specific range $\big(l, u\big)$; hence the pulse generator must be defined in this range. Given a system $S: U \rightarrow X$ that maps an input signal $u\in U$ to an output signal $x\in X$, and a specification $\varphi$ for $x$, the falsification problem is the problem of finding $u$ such that $x$ violates $\varphi$. The usual approach to solve this problem consists in defining a quantitative semantics $\rho(x)$ for $\varphi$ such that if $\rho(x)<0$ then $x$ violates $\varphi$, defining a parameterization $p \rightarrow u(p)$, and minimizing $\rho$ over some range of $p$ until it becomes negative. In this paper, we consider the special case where inputs are defined as periodic square waves and evaluate their ability to solve a suite of existing falsification benchmark problems.

\subsection{Pulse Generator Parameterization}

\begin{figure}
    \centering
    \includegraphics[height=6.7cm]{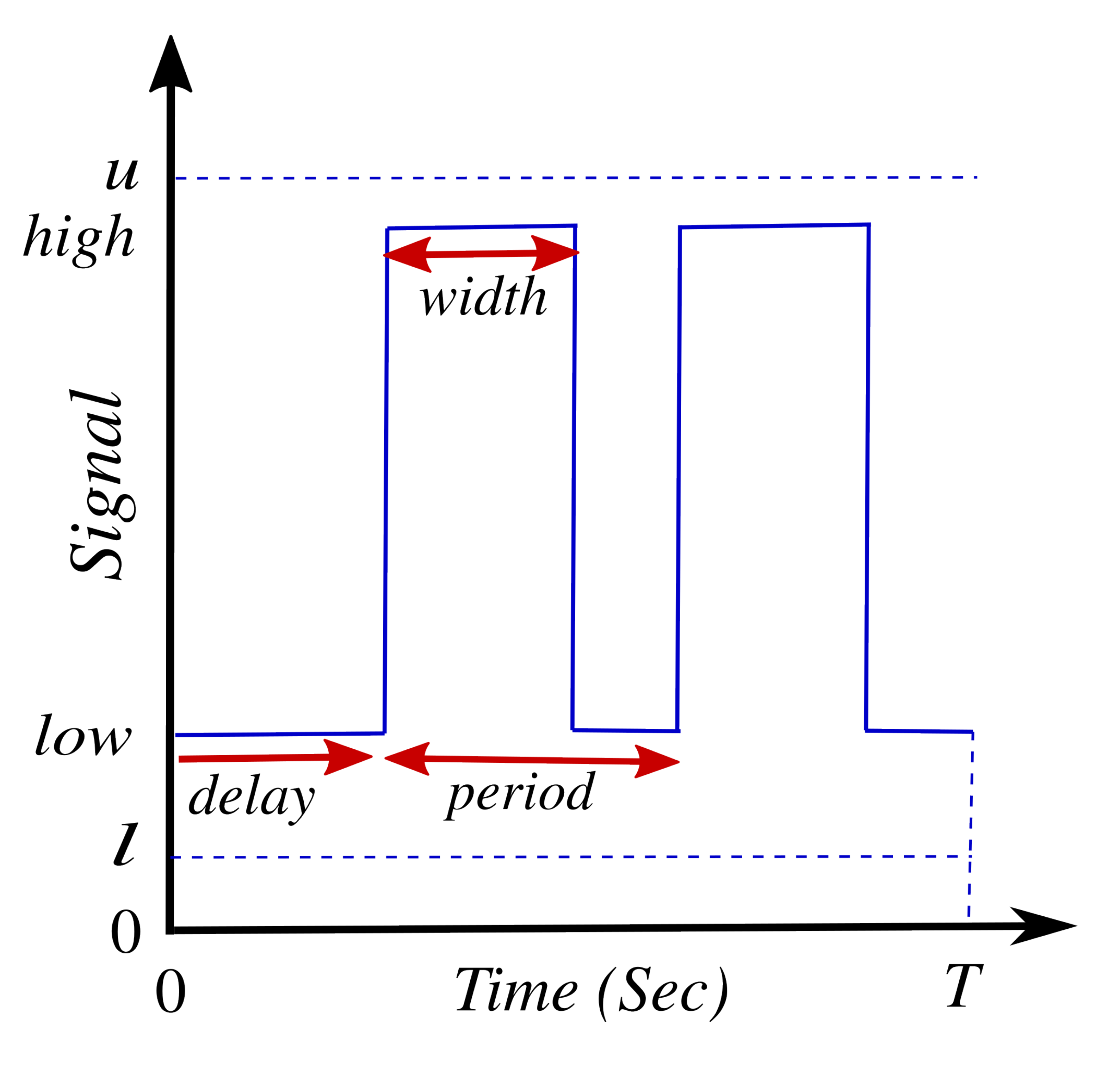}
    \caption{Pulse generator parameterization: A pulse generator can be built with $ (\periodPrime, \widthPrime, \highPrime, \lowPrime, \delayPrime)$}
    \label{fig:Pulse}
\end{figure}

Periodic square wave pulses are shown in Figure~\ref{fig:Pulse} which is parameterized differently from our work in~\cite{Ramezani2021temporal}. A pulse generator can be defined by five parameters $p = (\periodPrime$, $\widthPrime$, $\highPrime$, $\lowPrime$, $\delayPrime)$ where $T$ is the simulation time. 
If $\period$ and $\delay$ are relatively small wrt $T$, we get regular square-shaped inputs. However, we can also obtain signals of different types, e.g.,
\begin{itemize}
    \item {\bf Constant signals}: $\periodPrime \geq 2$, $\widthPrime \geq0.5$, and $\delayPrime > 0$
    \item {\bf Single step inputs}: $\periodPrime = 2 $ and $\delayPrime$ varies in $[0, 1]$.
\end{itemize}

To generate the pulse generator presented in Fig.~\ref{fig:Pulse}, we parameterized it as follows:

\begin{gather*}
\label{eq:and-additive}
   \mathit{period} = period^\prime * T \\
   width = width^\prime * period \\
   delay = delay^\prime * T \\
   low = l + low^\prime * (u - l) \\
   high = low + high^\prime * (u - low) \\ 
\end{gather*}

The parameters $width^\prime, delay^\prime, low^\prime, high^\prime$ are in $[0~1]$. $period^\prime$ can vary in two different ranges depending on the use of $delay^\prime$ as the input pulse generator. 
If $period^\prime$ and $delay^\prime$ are used together, $period^\prime = [0~1]$; otherwise $period^\prime = [0~2]$. This assumption is used because if $delay^\prime= 1$, it can make a constant signal equal to $low$. On the other hand, when the $delay^\prime$ is not included, $period^\prime = 2$, can make a step signal.

\section{Experimental Evaluation}
\label{Sec:Evaluation-Results}
The pulse generator is evaluated on the ARCH benchmark examples of~\cite{Ernst2019-ARCH19:} and~\cite{Ramezani2020-wodes}. These examples are:

\begin{itemize}
    \renewcommand{\labelitemi}{$\bullet$}
    \item Automatic Transmission (\AT): Two inputs, throttle in [0, 100], and brake in [0, 325];
    \item Automatic Transmission ($AT^\prime$): Same as \AT, but with different specifications and the brake input in [0, 500];
    \item Chasing Cars (\CC): Two inputs, throttle, and brake, both defined in [0, 1];
    \item $\Delta - \Sigma$ Modulator: Single input, three different ranges: [-0.35, 0.35], [-0.40, 0.40], [-0.45, 0.45]; with three initial conditions $x_{1}^{init}$, $x_{2}^{init}$, $x_{3}^{init}$ in the range [-0.1, 0.1];
    \item Switched System (\switchedS): Two inputs defined in the range [-1, 1]; Three different values are considered for parameter $thresh$: 0.7, 0.8, 0.9;
    \item Neural Network (\NN): One input (reference) chosen in [1, 3];
    \item Fuel Control (\AFC): Two inputs, throttle in [0, 61.2], and engine speed in [900, 1000];
    \item Steam Condenser (\SC): One input with possible values in [3.99, 4.01].
\end{itemize}

In the following, we refer to all these examples as ARCH benchmark examples.

\subsection{Evaluation on dimensioanlity}

Based on the evaluation result, it is hard to decide how many input parameters are needed to falsify a specification because it heavily depends on the \SUT. To show this fact, an experimental setup is given in this section. For this purpose, the pulse generator is applied to the benchmark examples. The pulse generator has five parameters to be built, $\lowPrime$ (L)), $\periodPrime$ (P), $\widthPrime$ (W), $\delayPrime$ (D), $\highPrime$ (H). Hence, first, only one input parameter for the optimization problem is considered, while the other input parameters which are not included in the optimization problem are constant. These values are: $\lowPrime = 0$, $\periodPrime = 0.5$, $\widthPrime = 0.5$, $\delayPrime = 0$,  and $\highPrime = 1$. 
The parameters $\lowPrime$, $\widthPrime$, $\delayPrime$, $\highPrime$, and $\highPrime$ can vary in $[0~1]$ when each is included as input for the falsification process. On the other hand, $\periodPrime$ can vary in two different ranges depending on the use of $\delayPrime$ as the input pulse generator. If $\periodPrime$ and $\delayPrime$ are used together, $\periodPrime = [0~1]$; otherwise $\periodPrime = [0~2]$. This assumption is used because if $\delayPrime=1$, it can make a constant signal equal to the lower bound of the pulse signal. On the other hand, when the $\delayPrime$ is not included, $\periodPrime = 2$ can make a step signal. 

We evaluate pulse generator inputs using a Bayesian optimization method (\BO)~\cite{Shahriari2016}, called \turbo~\cite{Eriksson2019turbo}. \turbo is a  trust-region BO method that locally exploits the objective function with a local optimization runs sequence. \turbo shows a good performance for falsification of \CPSs in~\cite{Ramezani2022Bayesian}. As \turbo requires a set of initial samples to start the process, we set the initial number of samples to $2\cdot n$, where $n$ is the number of input parameters that shows the dimensionality of the optimization problem. As \turbo requires a set of initial samples to start the process, we set the initial number of samples to $2\cdot$  number of input parameters.

\begin{table*}[!htbp] \scriptsize
\renewcommand{\arraystretch}{1.5}
\caption{An evaluation results base on one assessment to show how many and which input parameters are needed to falsify the evaluated specifications. The maximum number of the evaluated specifications is 40.}
\label{tab:Number-Inputs}
\centering
\input{Tables/Pulse_Turbo_TS}
\end{table*}

The results for the ARCH benchmark examples with only one input parameter are presented in Table~\ref{tab:Easy_Examples}-~\ref{tab:Special_Cases}, in the second to the sixth column. In these tables, the first column denotes the specifications; and the rest of the columns contain the evaluation results in which input parameters are included for the optimization, starting from single inputs to including all input parameters.
Each falsification is set to have the maximum number of simulations as $N$ = 1000. There are 5 independent falsification repetitions for each method and objective function to account for most algorithms' random nature.
Two values are presented for each specification; the first is the relative success rate of falsification in percent. There are 20 falsification runs for each parameter value and specification; thus, the success rate will be a multiple of 20\%. The second value, inside parentheses, is the average number of simulations (rounded) \emph{per successful falsification}. 

Table~\ref{tab:Number-Inputs} demonstrates evaluation results to show which combination of input generators might work better. 
To do this assessment, we evaluate if at least one input parameter is successful in falsifying a specification; regardless of the success rate, their combination also will be successful in falsifying it. For example, for the $\varphi_1^{AT^\prime} (T = 2)$ in Table~\ref{tab:Adding_More_Inputs}, while the three input $\lowPrime - \periodPrime -\widthPrime$ are successful to falsify, the $\highPrime$ and $\delayPrime$ are not. Hence, we assume that all combinations of different input parameters can falsify this except the combination $\highPrime - \delayPrime$. Based on this evaluation, in Table~\ref{tab:Number-Inputs}, the first row shows when we have one to five input parameters for the optimization, which combination of them gives the highest chance of falsifying a specification. The second row shows how many specifications out of 40 evaluated specifications are falsified when a different number of inputs are included in the optimization problem. Based on this evaluation, we can see that the best input parameter when one input is allowed to vary is $\widthPrime$, which falsifies most specifications (30 out of 40). While the three combinations $\lowPrime - \periodPrime$, $\lowPrime - \widthPrime$, and $\periodPrime - \widthPrime$ were expected to provide the best performance, the combination $\lowPrime - \widthPrime$ is more successful when only two parameters are allowed. We have only one option for three input parameters $\lowPrime - \periodPrime - \widthPrime$, which shows the best performance. As can be seen that four input parameters are the best results with falsifying all specifications at least in one run. Contrary, when all pulse generator parameters are included, 38 specifications are falsified. 
Hence, the rest of the columns of tables~\ref{tab:Easy_Examples}-~\ref{tab:Special_Cases} present the results for which combination was expected to give better results. 

\subsection{Evaluation Results}
\label{Sec:experimenta}

The results for the ARCH benchmark examples are easily falsified with a few simulations regardless of how many dimensions are presented in Table~\ref{tab:Easy_Examples}. Table~\ref{tab:Adding_More_Inputs} shows the results for some examples are falsified if more pulse input parameters are assumed for the optimizations. On the other hand, those examples that if we add more input parameters make the optimization hard to solve, which means that make the specifications hard to falsify, are presented in Table~\ref{tab:Adding_More_Inputs_DoNot_Help}. 
Additionally, we also include the results for the specifications that depend on which input parameters are added in Table~\ref{tab:Special_Cases}, sometimes having more input parameters helps but sometimes does not help to falsify regardless of the optimization methods.

\begin{table*}[!htbp] \tiny
\renewcommand{\arraystretch}{1.5}
\caption{Results for easy specifications are always falsifiable regardless of the number of input parameters using the pulse generator.}
\label{tab:Easy_Examples}
\centering
\input{Tables/Easy_Examples}
\end{table*}

\begin{table*}[!htbp] \tiny
\renewcommand{\arraystretch}{1.5}
\caption{The results for the specifications that if more input parameters are added, i.e., increasing the dimensionality of the optimization problem, can lead to falsifying them using the pulse generator.}
\label{tab:Adding_More_Inputs}
\centering
\input{Tables/Adding_More_Inputs}
\end{table*}

\begin{table*}[!htbp] \tiny
\renewcommand{\arraystretch}{1.5}
\caption{The results for the specifications that if more input parameters are added, i.e., increasing the dimensionality of the optimization problem, cannot lead to falsifying them using the pulse generator.}
\label{tab:Adding_More_Inputs_DoNot_Help}
\centering
\input{Tables/Adding_More_DoesNot_help}
\end{table*}

\begin{table*}[!htbp] \tiny
\renewcommand{\arraystretch}{1.5}
\caption{The results for the specifications that if more input parameters are added, i.e., increasing the dimensionality of the optimization problem, can or cannot lead to falsifying them using the pulse generator, depending on which input parameters are used in the pulse generator.}
\label{tab:Special_Cases}
\centering
\input{Tables/Special_Cases}
\end{table*}

As can be seen in Table~\ref{tab:Adding_More_Inputs}, increasing the number of dimensions improves the falsification performance. For example, while none of single inputs is successful to falsify $\varphi_6^{AT^\prime} (T = 10)$ and $(T = 12)$ in Table~\ref{tab:Adding_More_Inputs}, we can see a good performance with 100\% using two combination of $\lowPrime - \widthPrime$ and four inputs and also five inputs.
On the other hand, in Table~\ref{tab:Adding_More_Inputs_DoNot_Help}, we can see examples that can work with one single input. However, adding more inputs to that single input makes the specification harder to falsify or requires more simulations. For example, while $\widthPrime$ falsifies the $\varphi_4^{CC}$ easily, the success rate is decreased to zero when all inputs are included. Similarly, for the example \SC, while $\periodPrime$ works quite well, adding more inputs is not a useful option. 

Table~\ref{tab:Special_Cases} shows some special cases. While $\periodPrime$ is not successful to falsify $\varphi_1^{AT^\prime}$ with different $T$ values and $\varphi_2^{CC}$, the combination of $\periodPrime$ with $\lowPrime$ is not successful as well. As a result, it can be seen that for this specification, the effect of frequency is more than the pulse height domain. On the other hand, $\period$ is not an important input for $\varphi_8^{AT^\prime} (\bar{\omega} = 3000)$ and $\bar{\omega} = 3500$, and it is $\lowPrime$ that affect the result, as was done in~\cite{Ramezani2021temporal} where an inverted pulse was used to falsify them. $\highPrime$ only falsified one of these two examples in one run out of 5 that $\periodPrime$ or $\widthPrime$ did not do.

\begin{figure}
    \centering
    \includegraphics[height=9cm]{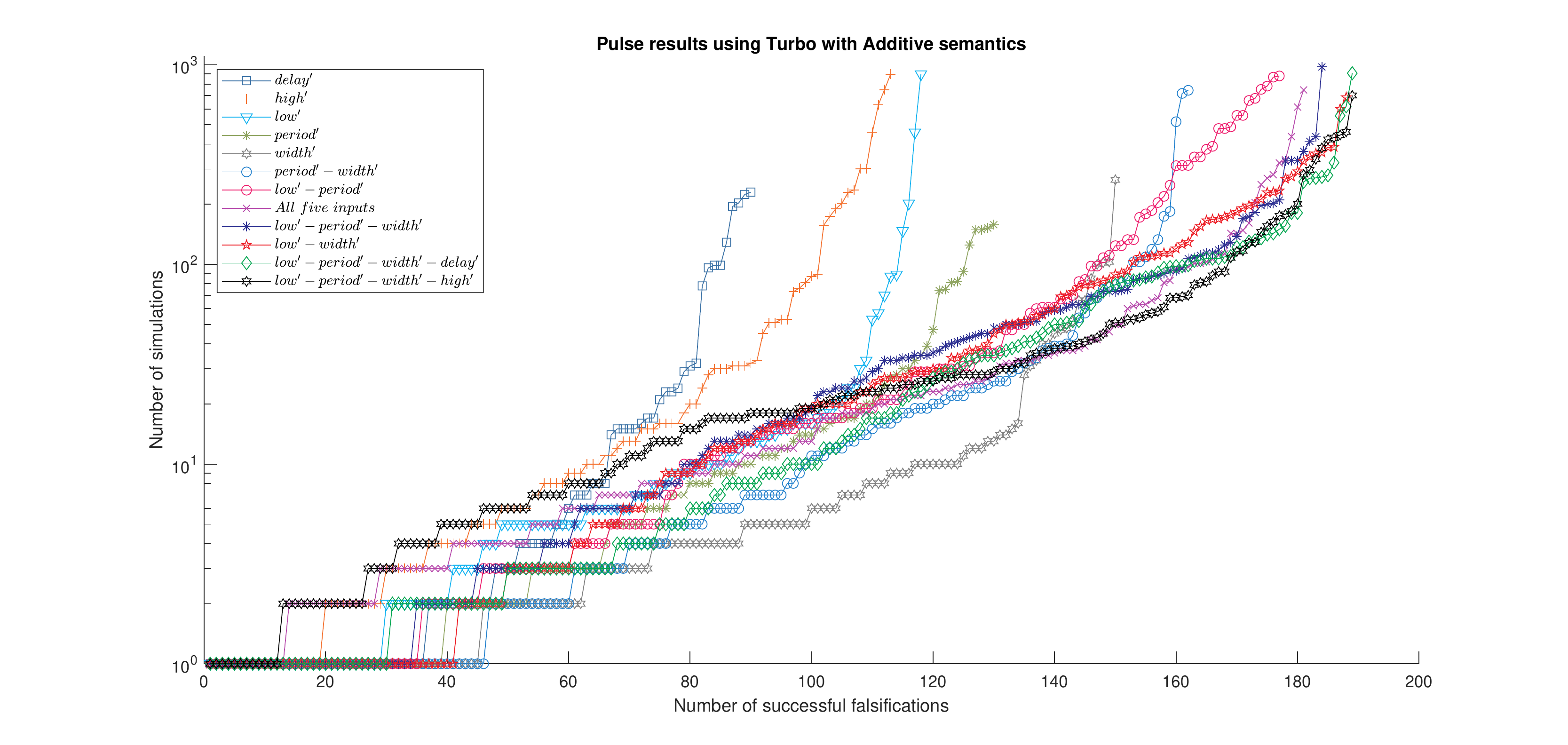}
    \caption{A cactus plot showing the results of different combinations of input parameters of a pulse generator using \turbo with \TS and \additiveS semantics.}
    \label{fig:Pulse-Results-Turbo}
\end{figure}

A cactus plot that shows the performance different
Fig.~\ref{fig:Pulse-Results-Turbo} shows an aggregated comparison including all evaluated methods. As the discussion obviously showed, $\widthPrime$ could falsify most of the exams with a few simulations. On the other hand, when more input parameters are added, generally, more specifications are falsified. Based on this evaluation. In general, $\delayPrime$ did not falsify any specifications that are not falsifiable with $\periodPrime$ or $\widthPrime$.
In a conclusion, as it is obvious from the evaluation, the combination of $\lowPrime - \widthPrime$ could falsify more specifications rather than when $\widthPrime$ combined with $\periodPrime$ and the combination of $\periodPrime - \highPrime$. All combinations of four inputs, when $\delayPrime$ or $\highPrime$ is not considered, show as well as good results. On the other hand, having all five parameters result in less falsification success than three, four, and the combination of $\lowPrime - \widthPrime$.

\section{Conclusion}
\label{Sec:Conclusion}

In this report, we evaluate the effect of dimensionality on the falsification performance using a pulse generator.
Our experiments demonstrate which input parameters are used or how many input parameters are needed for the optimization process, depending heavily on the application. Moreover, based on the results of this work, we could conclude that considering the input parameters in the time domain, i.e., $\periodPrime$ and $\widthPrime$, except $\delayPrime$ is more important than changes in the amplitude domain. In general, using the pulse generator results in falsifying all benchmark examples.

\section*{Acknowledgments}
This work was supported by the Swedish Research Council (VR) project
SyTeC VR 2016-06204 and from the Swedish Governmental Agency for
Innovation Systems (VINNOVA) under project TESTRON 2015-04893. The evaluations were performed using resources at High Performance Computing Center North (HPC2N), Umeå University, a Swedish national center for Scientific and Parallel Computing.

\bibliographystyle{unsrt}  
\bibliography{references.bib}

\end{document}

%% file: Tables/Pulse_Turbo_TS.tex
\begin{tabular}{c| c| c| c | c | c  c | c | c | c}

\hline

Number of Inputs
& \multicolumn{1}{c|}{One} 
& \multicolumn{1}{c|}{Two}
& \multicolumn{1}{c|}{Three} 
& \multicolumn{1}{c|}{Four}
& \multicolumn{1}{c}{Five} 
\\
\hhline{--------}

\hhline{--------}

Input  
& $\widthPrime$
& $\lowPrime - \periodPrime$  
& $\lowPrime - \periodPrime - \widthPrime$
& $\lowPrime - \periodPrime - \widthPrime - \highPrime$ 
& all five 

\\
combinations
& 
& $\lowPrime - \widthPrime$ 
& 
& $\lowPrime - \periodPrime - \widthPrime - \delayPrime$
& parameters
\\

& 
& $\periodPrime - \widthPrime$ 
& 
&
& 
\\

\hhline{--------}
Num. of falsified spec.
& 30
& 33
& 38
& 40
& 38 
\\

\hhline{--------}
\hhline{--------}

\end{tabular}

%% file: Tables/Easy_Examples.tex
\begin{tabular}{c c c c c c c c c c c c c c c c }

\hline

Spec
& \multicolumn{1}{c}{L} 
& \multicolumn{1}{c}{P}
& \multicolumn{1}{c}{W} 
& \multicolumn{1}{c}{H}
& \multicolumn{1}{c}{D}
& \multicolumn{1}{c}{L-P}
& \multicolumn{1}{c}{L-W}
& \multicolumn{1}{c}{P-W}
& \multicolumn{1}{c}{L-P-W}
& \multicolumn{1}{c}{L-P-W-H}
& \multicolumn{1}{c}{L-P-W-D}
& \multicolumn{1}{c}{L-P-W-H-D}
\\

\toprule

\multirow{3}{*}{}

$\varphi_2^{AT}$
& \textbf{100} (6)
& \textbf{100} (2)
& \textbf{100} (2)
& \textbf{100} (7)
& \textbf{100} (5)
& \textbf{100} (2)
& \textbf{100} (1)
& \textbf{100} (2)
& \textbf{100} (2)
& \textbf{100} (5)
& \textbf{100} (2)
& \textbf{100} (7)
\\

$\varphi_5^{AT}$
& \textbf{100} (1)
& \textbf{100} (3)
& \textbf{100} (2)
& \textbf{100} (3)
& \textbf{100} (1)
& \textbf{100} (3)
& \textbf{100} (3)
& \textbf{100} (5)
& \textbf{100} (4)
& \textbf{100} (13)
& \textbf{100} (3)
& \textbf{100} (8)
\\

\midrule

$\varphi_4^{AT^\prime} (T = 2)$
& \textbf{100} (3)
& \textbf{100} (11)
& \textbf{100} (5)
& \textbf{100} (2)
& \textbf{100} (3)
& \textbf{100} (81)
& \textbf{100} (10)
& \textbf{100} (8)
& \textbf{100} (22)
& \textbf{100} (16)
& \textbf{100} (7)
& \textbf{100} (18)
\\

$\varphi_5^{AT^\prime} (T = 2)$
& \textbf{100} (1)
& \textbf{100} (3)
& \textbf{100} (3)
& \textbf{100} (21)
& \textbf{100} (23)
& \textbf{100} (2)
& \textbf{100} (2)
& \textbf{100} (1)
& \textbf{100} (1)
& \textbf{100} (7)
& \textbf{100} (1)
& \textbf{100} (7)
\\

\midrule

$\varphi_1^{\Delta - \Sigma} ([−0.35, 0.35])$
& \textbf{100} (57)
& \textbf{100} (15)
& \textbf{100} (33)
& \textbf{100} (221)
& \textbf{100} (58)
& \textbf{100} (28)
& \textbf{100} (128)
& \textbf{100} (9)
& \textbf{100} (39)
& \textbf{100} (85)
& \textbf{100} (49)
& \textbf{100} (97)
\\

$\varphi_1^{\Delta - \Sigma} ([−0.40, 0.40])$
& \textbf{100} (12)
& \textbf{100} (5)
& \textbf{100} (7)
& \textbf{100} (12)
& \textbf{100} (10)
& \textbf{100} (31)
& \textbf{100} (35)
& \textbf{100} (8)
& \textbf{100} (15)
& \textbf{100} (24)
& \textbf{100} (20)
& \textbf{100} (39)
\\

$\varphi_1^{\Delta - \Sigma} ([−0.45, 0.45])$
& \textbf{100} (18)
& \textbf{100} (6)
& \textbf{100} (2)
& \textbf{100} (12)
& \textbf{100} (9)
& \textbf{100} (25)
& \textbf{100} (56)
& \textbf{100} (24)
& \textbf{100} (18)
& \textbf{100} (32)
& \textbf{100} (55)
& \textbf{100} (18)
\\

\midrule

$\varphi_1^{SS} (\gamma = 0.7)$
& \textbf{100} (1)
& \textbf{100} (1)
& \textbf{100} (1)
& \textbf{100} (48)
& \textbf{100} (1)
& \textbf{100} (1)
& \textbf{100} (1)
& \textbf{100} (1)
& \textbf{100} (1)
& \textbf{100} (9)
& \textbf{100} (1)
& \textbf{100} (4)
\\

\midrule

$\varphi_1^{AFC}$
& \textbf{100} (6)
& \textbf{100} (1)
& \textbf{100} (1)
& \textbf{100} (3)
& \textbf{100} (1)
& \textbf{100} (16)
& \textbf{100} (3)
& \textbf{100} (3)
& \textbf{100} (1)
& \textbf{100} (17)
& \textbf{100} (113)
& \textbf{100} (28)
\\

$\varphi_2^{AFC} $
& \textbf{100} (1)
& \textbf{100} (1)
& \textbf{100} (1)
& \textbf{100} (1)
& \textbf{100} (1)
& \textbf{100} (11)
& \textbf{100} (1)
& \textbf{100} (1)
& \textbf{100} (1)
& \textbf{100} (1)
& \textbf{100} (2)
& \textbf{100} (1)
\\

\midrule

$\varphi_1^{NN}$ 
& \textbf{100} (12)
& \textbf{100} (4)
& \textbf{100} (4)
& \textbf{100} (62)
& \textbf{100} (19)
& \textbf{100} (10)
& \textbf{100} (10)
& \textbf{100} (3)
& \textbf{100} (7)
& \textbf{100} (8)
& \textbf{100} (7)
& \textbf{100} (5)
\\

\bottomrule

\end{tabular}

%% file: Tables/Adding_More_Inputs.tex
\begin{tabular}{c c c c c c c c c c c c c c c c }

\hline

Spec
& \multicolumn{1}{c}{L} 
& \multicolumn{1}{c}{P}
& \multicolumn{1}{c}{W} 
& \multicolumn{1}{c}{H}
& \multicolumn{1}{c}{D}
& \multicolumn{1}{c}{L-P}
& \multicolumn{1}{c}{L-W}
& \multicolumn{1}{c}{P-W}
& \multicolumn{1}{c}{L-P-W}
& \multicolumn{1}{c}{L-P-W-H}
& \multicolumn{1}{c}{L-P-W-D}
& \multicolumn{1}{c}{L-P-W-H-D}
\\

\toprule

\multirow{3}{*}{}

$\varphi_2^{AT^\prime} (T = 10)$

& \textbf{100} (9)
& \textbf{100} (2)
& \textbf{100} (3)
& 0 (-)
& 0 (-)
& \textbf{100} (2)
& \textbf{100} (3)
& \textbf{100} (3)
& \textbf{100} (2)
& \textbf{100} (2)
& \textbf{100} (4)
& \textbf{100} (3)

\\

$\varphi_4^{AT^\prime} (T = 1)$

& 0 (-)
& \textbf{100} (14)
& \textbf{100} (10)
& \textbf{100} (3)
& \textbf{100} (3)
& \textbf{100} (30)
& \textbf{100} (34)
& \textbf{100} (6)
& \textbf{100} (12)
& \textbf{100} (11)
& \textbf{100} (11)
& \textbf{100} (7)
\\

$\varphi_5^{AT^\prime} (T = 1)$

& \textbf{100} (6)
& \textbf{100} (2)
& \textbf{100} (2)
& 0 (-)
& 0 (-)
& \textbf{100} (2)
& \textbf{100} (1)
& \textbf{100} (2)
& \textbf{100} (3)
& \textbf{100} (11)
& \textbf{100} (2)
& \textbf{100} (10)
\\

$\varphi_6^{AT^\prime} (T = 10)$

& 0 (-)
& 0 (-)
& 0 (-)
& 0 (-)
& 0 (-)
& 0 (-)
& \textbf{100} (230)
& 0 (-)
& 60 (184)
& 80 (69) 
& \textbf{100} (89)
& \textbf{100} (58) 

\\

$\varphi_6^{AT^\prime} (T = 12)$

& 0 (-)
& 0 (-)
& 0 (-)
& 0 (-)
& 0 (-)
& 60 (551)
& \textbf{100} (174)
& 80 (509) 
& \textbf{100} (279)
& \textbf{100} (24) 
& \textbf{100} (49) 
& \textbf{100} (16)
\\

$\varphi_7^{AT^\prime}$

& 0 (-)
& 0 (-)
& 0 (-)
& 0 (-)
& 0 (-)
& 0 (-)
& 0 (-)
& 0 (-)
& 0 (-)
& 80 (221)
& \textbf{100} (96)
& 80 (250) 
\\
\midrule

$\varphi_1^{AT}$ 
& 0 (-)
& 0 (-)
& \textbf{100} (80)
& 0 (-)
& 0 (-)
& 0 (-)
& \textbf{100} (81)
& \textbf{100} (21)
& \textbf{100} (75)
& \textbf{100} (79)
& \textbf{100} (78)
& \textbf{100} (60)
\\

$\varphi_3^{AT}$
& 0 (-)
& \textbf{100} (123)
& \textbf{100} (7)
& 0 (-)
& 0 (-)
& \textbf{100} (167)
& \textbf{100} (29)
& \textbf{100} (28)
& \textbf{100} (102)
& \textbf{100} (238)
& \textbf{100} (9)
& \textbf{100} (40)
\\

$\varphi_4^{AT}$
& 0 (-)
& \textbf{100} (2)
& \textbf{100} (5)
& 0 (-)
& 0 (-)
& \textbf{100} (20)
& \textbf{100} (6)
& \textbf{100} (8)
& \textbf{100} (20)
& \textbf{100} (7)
& \textbf{100} (6)
& \textbf{100} (16)
\\

\midrule

$\varphi_1^{CC}$ 
& \textbf{100} (3)
& \textbf{100} (21)
& \textbf{100} (3)
& \textbf{100} (6)
& 0 (-)
& \textbf{100} (34)
& \textbf{100} (15)
& \textbf{100} (4)
& \textbf{100} (17)
& \textbf{100} (13)
& \textbf{100} (4)
& \textbf{100} (7)
\\

$\varphi_3^{CC}$
& \textbf{100} (8)
& \textbf{100} (16)
& \textbf{100} (8)
& \textbf{100} (30)
& 0 (-)
& \textbf{100} (132)
& \textbf{100} (42)
& \textbf{100} (31)
& \textbf{100} (62)
& \textbf{100} (26)
& \textbf{100} (21)
& \textbf{100} (15)
\\

$\varphi_5^{CC}$
& 0 (-)
& \textbf{100} (14)
& \textbf{100} (16)
& \textbf{100} (19)
& 0 (-)
& \textbf{100} (345)
& \textbf{100} (13)
& \textbf{100} (19)
& \textbf{100} (17)
& \textbf{100} (27)
& \textbf{100} (64)
& \textbf{100} (8)
\\

\midrule

$\varphi_1^{SS} (\gamma = 0.8)$
& \textbf{100} (1)
& \textbf{100} (1)
& \textbf{100} (1)
& 60 (280)
& \textbf{100} (1)
& \textbf{100} (1)
& \textbf{100} (1)
& \textbf{100} (1)
& \textbf{100} (1)
& \textbf{100} (4)
& \textbf{100} (11)
& \textbf{100} (170)
\\

$\varphi_1^{SS} (\gamma = 0.9)$
& \textbf{100} (1)
& \textbf{100} (1)
& \textbf{100} (1)
& 40 (461)
& \textbf{100} (1)
& \textbf{100} (1)
& \textbf{100} (1)
& \textbf{100} (1)
& \textbf{100} (1)
& \textbf{100} (41)
& \textbf{100} (1)
& \textbf{100} (11)
\\

\midrule
$\varphi_2^{NN}$
& 60 (519)
& \textbf{100} (11)
& \textbf{100} (5)
& 80 (520)
& \textbf{100} (190)
& \textbf{100} (8)
& \textbf{100} (8)
& \textbf{100} (11)
& \textbf{100} (53)
& \textbf{100} (11)
& \textbf{100} (7)
& \textbf{100} (14)
\\
\bottomrule

\end{tabular}

%% file: Tables/Adding_More_DoesNot_help.tex
\begin{tabular}{c c c c c c c c c c c c c c c c }

\hline

Spec
& \multicolumn{1}{c}{L} 
& \multicolumn{1}{c}{P}
& \multicolumn{1}{c}{W} 
& \multicolumn{1}{c}{H}
& \multicolumn{1}{c}{D}
& \multicolumn{1}{c}{L-P}
& \multicolumn{1}{c}{L-W}
& \multicolumn{1}{c}{P-W}
& \multicolumn{1}{c}{L-P-W}
& \multicolumn{1}{c}{L-P-W-H}
& \multicolumn{1}{c}{L-P-W-D}
& \multicolumn{1}{c}{L-P-W-H-D}
\\

\toprule

$\varphi_3^{AT^\prime} (T = 4.5)$ 

& 0 (-)
& \textbf{100} (60)
& \textbf{100} (6)
& 0 (-)
& 0 (-)
& \textbf{100} (358)
& \textbf{100} (104)
& \textbf{100} (94)
& 80 (387) 
& 20 (282)
& 20 (555) 
& 20 (747)
\\
\hhline{----------------}

$\varphi_3^{AT^\prime} (T = 5)$

& 0 (-)
& \textbf{100} (11)
& \textbf{100} (3)
& 0 (-)
& 0 (-)
& \textbf{100} (14)
& \textbf{100} (22)
& \textbf{100} (36)
& \textbf{100} (144)
& \textbf{100} (149)
& 80 (475) 
& 0 (-)
\\

\hhline{----------------}
$\varphi_6^{AT}$ 
& 0 (-)
& \textbf{100} (4)
& 0 (-)
& 0 (-)
& 0 (-)
& \textbf{100} (32)
& \textbf{100} (188)
& \textbf{100} (18)
& \textbf{100} (11)
& \textbf{100} (22)
& \textbf{100} (5)
& \textbf{100} (47)
\\

$\varphi_7^{AT}$
& 0 (-)
& 0 (-)
& 0 (-)
& 0 (-)
& 0 (-)
& 0 (-)
& \textbf{100} (243) 
& 0 (-)
& \textbf{100} (68)
& \textbf{100} (27)
& \textbf{100} (134)
& \textbf{100} (54)
\\

$\varphi_8^{AT}$
& 0 (-)
& 0 (-)
& 0 (-)
& 0 (-)
& 0 (-)
& 0 (-)
& 80 (152) 
& 0 (-)
& \textbf{100} (93) 
& \textbf{100} (35)
& \textbf{100} (93)
& \textbf{100} (32)
\\

$\varphi_9^{AT}$
& 0 (-)
& 0 (-)
& 0 (-)
& 0 (-)
& 0 (-)
& 0 (-)
& \textbf{100} (67) 
& 0 (-)
& \textbf{100} (60) 
& \textbf{100} (22)
& \textbf{100} (78)
& \textbf{100} (19) \\

\midrule

$\varphi_4^{CC}$
& 0 (-)
& 0 (-)
& \textbf{100} (97)
& 0 (-)
& 0 (-)
& 0 (-)
& 80 (218)
& \textbf{100} (66)
& 40 (99)
& 60 (84)
& 20 (122)
& 0 (-)
\\

\midrule

$\varphi_1^{SC}$
& 0 (-)
& \textbf{100} (70)
& 0 (-)
& 0 (-)
& 0 (-)
& 80 (686)
& 0 (-)
& 60 (113)
& 0 (-)
& 40 (426)
& 60 (287)
& 20 (251)
\\

\bottomrule

\end{tabular}

%% file: Tables/Special_Cases.tex
\begin{tabular}{c c c c c c c c c c c c c c c c }

\hline

Spec
& \multicolumn{1}{c}{L} 
& \multicolumn{1}{c}{P}
& \multicolumn{1}{c}{W} 
& \multicolumn{1}{c}{H}
& \multicolumn{1}{c}{D}
& \multicolumn{1}{c}{L-P}
& \multicolumn{1}{c}{L-W}
& \multicolumn{1}{c}{P-W}
& \multicolumn{1}{c}{L-P-W}
& \multicolumn{1}{c}{L-P-W-H}
& \multicolumn{1}{c}{L-P-W-D}
& \multicolumn{1}{c}{L-P-W-H-D}
\\

\toprule

\multirow{3}{*}{}

\multirow{3}{*}{}
$\varphi_1^{AT^\prime} (T = 20)$ 
& \textbf{100} (6)
& 0 (-)
& \textbf{100} (7)
& \textbf{100} (2)
& \textbf{100} (2)
& 0 (-)
& \textbf{100} (29)
& \textbf{100} (25)
& \textbf{100} (40)
& \textbf{100} (16)
& \textbf{100} (7)
& \textbf{100} (4)
\\

\hhline{----------------}
$\varphi_1^{AT^\prime} (T = 30)$ 

& \textbf{100} (10)
& 0 (-)
& \textbf{100} (10)
& \textbf{100} (2)
& \textbf{100} (3)
& 0 (-)
& \textbf{100} (36)
& \textbf{100} (25)
& \textbf{100} (34)
& \textbf{100} (13)
& \textbf{100} (16)
& \textbf{100} (4)
\\
\hhline{----------------}
$\varphi_1^{AT^\prime} (T = 40)$

& \textbf{100} (14)
& 0 (-)
& \textbf{100} (14)
& \textbf{100} (6)
& \textbf{100} (32)
& 0 (-)
& \textbf{100} (29)
& \textbf{100} (24)
& \textbf{100} (31)
& \textbf{100} (19)
& \textbf{100} (41)
& \textbf{100} (14)
\\

$\varphi_8^{AT^\prime} (\bar{\omega} = 3000)$

& \textbf{100} (16)
& 0 (-)
& 0 (-)
& 20 (15)
& 0 (-)
& \textbf{100} (23)
& \textbf{100} (26)
& 0 (-)
& \textbf{100} (20)
& \textbf{100} (97)
& \textbf{100} (30) 
& \textbf{100} (24)
\\

$\varphi_8^{AT^\prime} (\bar{\omega} = 3500)$

& \textbf{100} (50)
& 0 (-)
& 0 (-)
& 0 (-)
& 0 (-)
& \textbf{100} (16)
& \textbf{100} (39)
& 0 (-)
& \textbf{100} (70)
& \textbf{100} (86)
& \textbf{100} (157)
& \textbf{100} (212)
\\

\midrule

$\varphi_2^{CC}$
& \textbf{100} (4)
& 0 (-)
& \textbf{100} (4)
& \textbf{100} (6)
& 0 (-)
& 0 (-)
& \textbf{100} (254)
& \textbf{100} (5)
& \textbf{100} (51)
& \textbf{100} (13)
& \textbf{100} (3)
& \textbf{100} (3)
\\

\bottomrule

\end{tabular}